# Hysteresis of static polarizability of a ballistic ring in presence of intense high-frequency field

E.M. Epshtein[1], G.M. Shmelev[2]

[1] Institute for Radio Engineering and Electronics of the Russian Academy of Sciences,
Vvedenskii sq., 1, Fryazino, Moscow District, 141190 Russia;
e-mail: eme253@ms.ire.rssi.ru

[2] Volgograd State Pedagogical University,
Lenin av., 27, Volgograd, 400131, Russia;
e-mail: shmelevgm@rambler.ru

**Abstract.** A ballistic ring in presence of collinear dc and high-frequency electric fields is considered. A possibility is demonstrated of the system switching between two states by changing the strength of dc and/or high-frequency field. It leads to hysteretic behavior of the ring dc dipole moment as a function of the dc electric field.

It was shown [1 − 5] that quasi-one-dimensional ballistic rings can display nonlinear electrodynamical properties. The cause is that electrons in ring can move only along the ring circumference, whereas tangential component of the force driving the electrons depends on the electron position in the ring by nonlinear way. Such "geometrical nonlinearity" leads to a number of nonlinear electrodynamical effects, such as nonlinear scattering, absorption and emission, as well as generation of higher harmonics and dc component [1 − 5].

In the present work, we consider a situation where two, namely, dc and high-frequency collinear electric fields affect the ballistic ring. A possibility will be demonstrated of the system switching between two states by changing the strength of dc and/or high-frequency field. Such a situation looks as a nonequilibrium phase transition driven by electric field.

Consider an electron in a quasi-one-dimensional ring, i.e., in an annular quantum well confined with two concentric circular potential barriers (or in a quantum wire closed as a ring). The ring width $d$ is much smaller than its radius $R$, so that the confinement takes place only across the well width, while the electron motion along the ring circumference is classical, but ballistic one, i.e. the electron moves without collisions (the ring radius is small in comparison with the electron mean free path but large in comparison with the electron de Broglie wavelength). The ring is placed in a uniform dc electric field $\mathbf{E}$ parallel to the ring plane (between the plates of a parallel-plate capacitor); the screening length is large in comparison with $d$ so that the a uniform field affects the electrons in the ring. A high-frequency electric field $\mathbf{F}(t) = \mathbf{F}_0 \cos \omega t$ induced by a linearly polarized electromagnetic wave that propagates along the normal to the ring plane affects the electrons, too. The frequency $\omega$ is assumed high in comparison with the electron collision frequency.

Since the electrons in the ring can move only along the ring circumference, the driving tangential force takes form $f(\phi, t) = (E + F(t))\sin \phi$, where the angular coordinate $\phi$ is counted off from the field direction.

The electron motion under action of the fields is described by a nonlinear equation

$$\frac{d^2\phi}{dt^2} + \frac{e}{mR}(E + F_0 \cos \omega t)\sin \phi = 0 \qquad (1)$$

where $e$ and $m$ are electron charge and effective mass, respectively.

Equation (1) coincides with the equation of motion of a pendulum with suspension point vibrating with frequency ω along vertical direction. A number of consequences follow immediately from that fact (see [6, 7]):

1) under

$$\omega \gg \Omega \equiv \sqrt{\frac{e|\mathbf{E}|}{mR}} < \gamma \equiv \frac{e|\mathbf{F}_0|}{\sqrt{2}mR\omega}$$ (2)

conditions, an additional stable equilibrium point $\phi = \pi$ appears (besides the usual one $\phi = 0$);

2) under $\omega = 2\Omega + \varepsilon$, $\varepsilon \ll \Omega$ conditions, parametric resonance occurs in $|\varepsilon| < \gamma\sqrt{2}$ range;

3) a route to a dynamical chaos regime is possible with suitable parameter choice.

Using the averaging method [6] under conditions (2), we obtain an equation of slow electron motion averaged over the high frequency ω

$$\frac{d^2\Phi}{dt^2} + \frac{eE}{mR}\sin\Phi + \gamma^2\sin\Phi\cos\Phi = 0 ,$$ (3)

where $\Phi$ is angular coordinate averaged over high frequency which describes the slow motion.

The equation (3) corresponds to the motion in an effective potential

$$U(\Phi) = -\frac{eE}{mR}\cos\Phi - \frac{1}{2}\gamma^2\cos^2\Phi .$$ (4)

Once more analogy may be used here: the dimensionless energy $U(\Phi)$ has the same angular dependence as the energy of a single-domain ferromagnet with uniaxial anisotropy in external magnetic field $H$ collinear to the anisotropy axis [8]:

$$U(\Phi) = -MH\cos\Phi - K\cos^2\Phi ,$$ (5)

where $M$ is saturation magnetization, $K$ is anisotropy energy density, $\Phi$ is angle between the magnetization and external magnetic field vectors.

The first term in the right-hand side of Eq. (4) corresponds to the Zeeman energy in an external magnetic field, while the second one describes the anisotropy energy. Therefore, electric behavior of the ballistic ring under changing dc electric field should be analogous to magnetic behavior of the ferromagnet under changing external magnetic field, including the hysteresis phenomenon.

The potential (4) as a function of $\Phi$ has two minima at $\Omega < \gamma$ and one minimum at $\Omega > \gamma$; at $\Omega = \gamma$ one of the minima disappears by transformation to an inflection point. At $\Omega < \gamma$ and $kT \ll \dfrac{e^2 F_0^2}{m\omega^2}$, an electron placed in one of the wells of the two-well potential cannot leave it even if the other well is more deep. The electron stays in that metastable state until the well disappears and the electron is to go to the rest well.

In absence of the dc electric field, the electrons in the ring distribute between two potential wells of the averaged potential induced by the high-frequency field. Turning on the dc field parallel to the high-frequency field leads to changing relative depth of the wells. At $E > \dfrac{eF_0^2}{2mR\omega^2} \equiv E_c$, only one of the wells remains, in which all $N$ electrons present in the ring get together. The ring dipole moment is $NeR$ in that state. Under decreasing dc field down to zero and further to $-E_c$, the well with electrons disappears and the electrons go to another well centered at the opposite end of the ring diameter. The ring dipole moment changes polarity abruptly and becomes equal to $-NeR$. Therefore, dependence of the ring dipole moment on dc electric field in presence of collinear intense high-frequency field takes form of rectangular hysteresis loop (see figure).

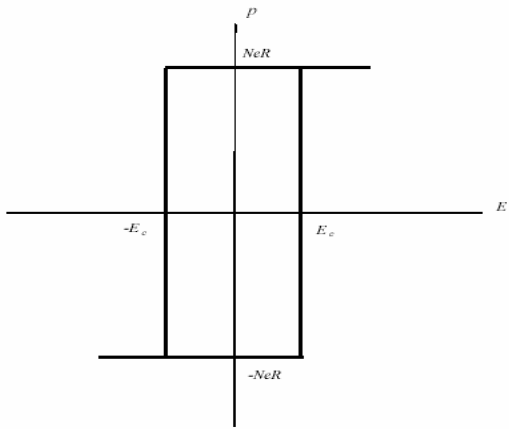

The present work was supported by Russian Foundation of Basic Research (grant RFBR No. 02-02-16238).

## References

[1] E.M. Epshtein, G.M. Shmelev. Physica Scripta **62**, 216 (2000).
[2] E.M. Epshtein, G.M. Shmelev, I.I. Maglevanny. J. Phys. A: Math. Gen. **33**, 6017 (2000).
[3] E.M. Epshtein, I.I. Maglevanny, G.M. Shmelev. Phys. Low-Dim. Struct. No. 3/4, 109 (2000).
[4] G.M.Shmelev, E.M. Epshtein, G.A. Syrodoev. Techn. Phys. **45**, 1354 (2000).
[5] E.M. Epshtein, G.M. Shmelev, I.I. Maglevanny. Phys. Low-Dim. Struct. No. 1/2, 137 (2001).
[6] L.D.Landau, E.M.Lifshitz. Mechanics (Pergamon, London, 1976).
[7] F. Moon. Chaotic Vibrations (Wiley, N.Y., 1987)].
[8] S. Chikazumi. Physics of Magnetism (Clarendon, Oxford, 1997).